# Extraction of Multi-layered Social Networks from Activity Data

Katarzyna Musial[1], Piotr Bródka[2], Przemysław Kazienko[2], Jarosław Gaworecki[3]
[1] School of Natural and Mathematical Sciences, Department of Informatics, King's College London, London, WC2R 2LS, United Kingdom
[2] Institute of Informatics, Wrocław University of Technology
Wyb.Wyspiańskiego 27, 50-370 Wrocław, Poland
[3] Research & Engineering Center Sp. z o.o., ul. Strzegomska 46B, 53-611 Wrocław, Poland
katarzyna.musial@kcl.ac.uk, piotr.brodka@pwr.wroc.pl, kazienko@pwr.wroc.pl, jaroslaw.gaworecki@rec-global.com

**Abstract**

*The data gathered in all kind of web-based systems, which enable users to interact with each other, provides an opportunity to extract social networks that consist of people and relationships between them. The emerging structures are very complex due to the number and type of discovered connections. In web-based systems, the characteristic element of each interaction between users is that there is always an object that serves as a communication medium. This can be e.g. an email sent from one user to another or post at the forum authored by one user and commented by others. Based on these objects and activities that users perform towards them, different kinds of relationships can be identified and extracted. Additional challenge arises from the fact that hierarchies can exist between objects, e.g. a forum consists of one or more groups of topics, and each of them contains topics that finally include posts. In this paper, we propose a new method for creation of multi-layered social network based on the data about users activities towards different types of objects between which the hierarchy exists. Due to the flattening, pre-processing procedure new layers and new relationships in the multi-layered social network can be identified and analysed.*

## 1. Introduction

Nowadays, for the first time, we have possibility to process big data about interactions and activities of millions of individuals gathered in all sort of web-based systems. Communication technologies allow us to form large networks, which in turn shape and catalyse our activities. Due to their scale, complexity and dynamics, these networks are extremely difficult (or impossible) to analyse in terms of traditional social network analysis methods. The analysis of the networked data is at the very early stages and requires a lot of effort in both developing tools and approaches to tackle it as well as understanding the nature and functioning of networks extracted from this data. The process of network creation is not as straightforward as it seems to be. In the web-based systems, users can interact with each other via different communication channels and utilize various services. This implies that the relationships between users can be extracted based on both direct and indirect communication. The former one is e.g. sending emails or video calls where information is passed directly from one person (group of people) to other(s), whereas to the latter we can count in e.g. commenting objects in the multimedia sharing systems or using the same tags to describe the objects. In both situations there are objects (e.g. email, photo, tag) that serve as medium in communication between users. The types of these objects differ depending on the web system, e.g. at the Internet forum the objects are: groups of topics, topics, posts while in the email service it will be a single message. Additionally, within a single system, the hierarchies of objects can exist.

Extraction of a social network in the environment where users interact with each other using different objects, which create hierarchies, is the main contribution of this paper. In order to perform this task first the hierarchical pre-social network (HPSN), where relation between users and objects as well as between objects exist (Section 5), must be created. After that the flattening process, in which the hierarchy of objects is removed, is performed (Section 6). As a result, the flat pre-social network (FPSN) is obtained where the only connections that exist are between users and one type of the previously chosen objects. Based on FPSN, social network, where only the connections between users exist, is created (Section 7). The whole idea is presented using simple case study of Internet forum (Section 8). Finally, the real-world experiments were performed and their outcomes are presented in Section 9.

## 2. Related Work

There are many types of complex networked systems. One of the classifications distinguishes infrastructures and natural complex systems [5]. The former are physical systems (energy and transportation networks) and virtual systems (Internet, WWW, telecommunication), whereas the latter are biological networks, social networks, food webs and ecosystems.

The type of complex systems that is investigated in this paper is a social network formed by people who interact with each other or take part in common activities. The concept of social network has been described by different researchers [2, 35, 36, 39, 40] and the definition that is commonly and widely used is that *a social network* is a finite set of individuals, who are the nodes of the network, and activities or relations between them, which are represented by edges of the network. A social network (SN) commonly represents the mutual communication and/or activity occurring between users as well as their direction, intensity, and profile [21].

It should be noted that, during analysis of social networks, researchers usually take into account only one activity type while in most cases many different types of relationships exist between users. The special type of social networks that allows the representation of many different activities is called a multi-layered social network [4, 9, 10, 11, 17, 23, 33, 34, 39], a.k.a. multi-dimensional network [6, 16, 42] or multiplex networks [1, 15, 20, 21, 27]. Even the same authors use different names for this kind of complex networks, compare e.g. [17] and [20]. Overall, due to high complexity, such networks are more difficult to be extracted and analysed than simple one-layered networks.

Sociologists and psychologist typically create questionnaires and perform interviews in order to collect data which allow them to create and analyse social networks. However, nowadays, the rapid development of the Internet and telecommunication together with the ease of gathering vast amount of data have created the possibility for IT systems to provide vast amount of information about users activities. As a result, researchers have now easy access to big datasets about people's activities ready to analyse. Social networks can be extracted from e.g. bibliographic data [19], blogs [3], photos sharing systems like Flickr [25], e-mail systems [38], telecommunication data [7, 26], social services like Twitter [22] or Facebook [18, 37], video sharing systems like YouTube [14], Wikipedia [13] and much more. Moreover, the whole separate systems were created only for the extraction, aggregation and visualization of social networks [30, 31].

Nevertheless, as mentioned before, only few scientists have focused their research interests at multi-layer social network extraction from activity data [6, 8, 15, 24, 27, 29, 33, 41, 42]. Moreover, no one has studied the hierarchy and relationships between objects in this data. The only hierarchical dependencies in the social networks that were analysed were associated with the hierarchy between users such as employee-employer, the employee-manager, etc. [28]. Thus, the analysis of hierarchy between objects presented in this paper is a new approach to extract multi-layer social networks from the activity data.

## 3. Object-based Relationships

In web-based social systems, there is always an object that plays a role of 'a middleman' in a relationship between two users (Fig. 1) [32]. In the case of direct communication, people send emails to each other or make videoconferences or phone calls via VoIP services. In those cases all participants are aware of the existing relationship. However, sometimes two users can be in a relationship but they do not maintain it actively and consciously, e.g. people who comment on the same blog or participate in the same conference. These types of common activities can result in indirect relationships. The roles of both users in an indirect relationship, towards the object can be either the same or different (Fig. 2).

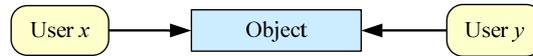

**Fig 1. The object-based relationship in the social network on the Internet**

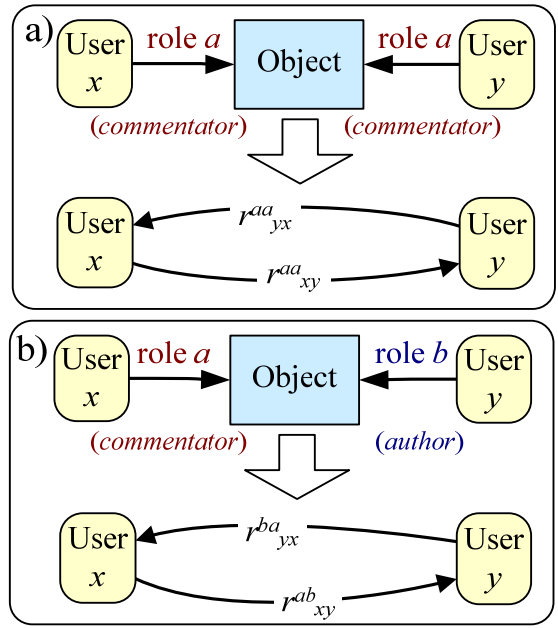

**Fig. 2. The object-based relation with equal roles:** *commentator* **(a), and different roles:** *commentator* **and** *author* **(b)**

*Object-based relation with equal roles* $r^{aa}_{xy}$ means that user *x* and *y* meet each other through the object and their role in relation to this object is the same. In other words, they participate in common activity related to the certain object with the same role *a*, e.g. two users take part in the videoconference, two users comment the same picture or both of them add the same object to their favourites [25], Fig. 2a. *Object-based relation with different roles* $r^{ab}_{xy}$, $r^{ba}_{yx}$ – is the relation between two users *x* and *y* that are connected through the object but their roles *a* and *b* towards the object are different, e.g. user *x* comments a photo (role *a* – commentator) that was published by *y* (role *b* – author) [25], Fig. 2b. The non-zero relation $r^{ab}_{xy}$ entails the non-zero relation $r^{ba}_{yx}$.

The examples of object-based relations with equal roles are:

(i) **Commentator-commentator** – this relation is created between user *x* and user *y* when both of them have added the opinions about at least one common object, e.g. they have commented the same picture at the photo sharing system or the same post at the forum.
(ii) **Favourite-favourite** – such a relation from user *x* to *y* exists if both users have marked at least one common object as their favourite, e.g. they have added the same film to their lists of favourites at the multimedia sharing system.
(iii) **Author-author** – such a relation from user *x* to *y* exists when they are co-authors of at least one object, e.g. they have written a scientific article together.
(iv) **Membership in the group/forum** – this relation from user *x* to *y* exists when both of them belong to at least one group together, e.g. they belong to a group that gathers people who like dogs at the photo sharing system.
(v) **Utilization of keywords to describe objects (tags)** – such a relation exists between two users if they use at least one common tag to describe their objects, e.g. two users are in relations with each other at the photo sharing system when they use a word "cat" to describe some of their photos.

On the other hand the examples of object-based relations with different roles can be:
(i) **Opinion-author and author-opinion** – these relations between user *x* and *y* exist when user *x* commented at least one object that is authored by user *y*.
(ii) **Favourite-author and author-favourite** – these relations between user *x* and *y* exist when user *x* added to its favourite list at least one object authored by user *y*.
(iii) **Citation-author and author-citation** – these relations between user *x* and *y* exist when user *x* quoted at least one object authored by user *y*.

## 4. Hierarchies Between Objects

In all web-based social networks analysed in the literature, the relationships between users were extracted mainly based on a given type of communication or common activity. For example, if two users send emails to each other, then the relationship between them in the social network may be established. However, both user communication and common activities are always related somehow to the objects which serve as a medium in interactions between users and their common activities, see section 3. This object may be 'a message' in the case of email exchange, 'a video' in YouTube or 'a topic' in the internet forums. These objects connect either a pair of users (an email sent to a single recipient) or many users simultaneously (an email passed to multiple recipients, a video commented by many users, a forum with many members). Besides, the IT system may provide many different functions, which can result in various user activities towards objects of different types. For example, an internet forum may consist of topics aggregated into groups. Topics, in turn, contain a list of posts. Thus, the objects that enable interactions between users are in hierarchical relationships, Fig. 3. Depending on the functionalities of the system, users can moderate a topic group, can subscribe to a topic (be a member of the topic) or provide their opinions about posts (play the role of commentator), Fig. 10. Hence, users 'meet' each other by performing activities towards objects that belong: (i) to one specific level in the object hierarchy (many users can comment a post authored by another user) or (ii) to two different levels of this hierarchy, e.g. a moderator of the group topic is in the indirect relation with authors of the posts.

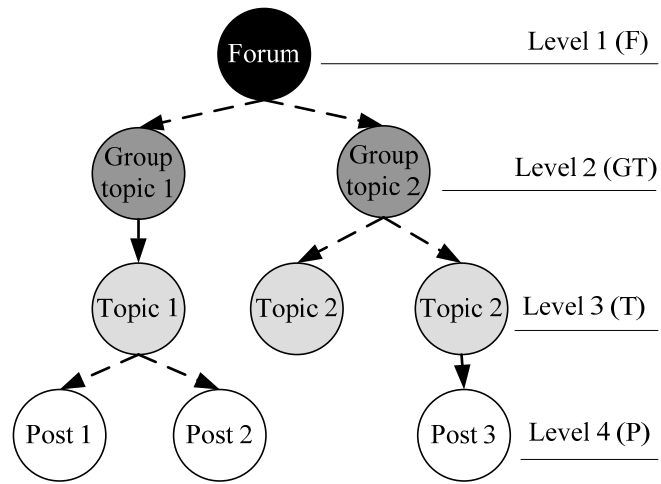

**Fig. 3 Hierarchies between objects in the Internet forum**

## 5. Hierarchical Pre-Social Network

In order to create the hierarchical pre-social network (HPSN) based on gathered activity data, first such elements as users, objects and hierarchy between objects as well as relations between users and objects need to be extracted. HPSN contains information about relations between users and objects towards which users performed some activities. The main characteristic of HPSN is that there exist hierarchies between different objects. The whole process of HPSN extraction consists of the following four steps:

**1) User extraction** – users are network nodes both in the pre-social network and the final social network. Users perform different activities towards various types of objects, e.g. they send emails to each other or comment the photos uploaded by others. These activities are the basis to create the role of a user in relation to a specific object, e.g. author, commentator, etc.

**2) Object extraction** – objects are the nodes in both hierarchical and flat pre-social network, i.e. elements through which users communicate with each other (e.g. email, phone call) or items towards which users perform some activities (e.g. photo, video, tag).

**3) Extraction of the hierarchy between the objects** – some objects can be in hierarchical relation with other objects, e.g. an object 'group of topics' contains one or more 'topics' which may include many 'posts' (Fig. 3). The consequence of the existence of the hierarchy between different objects types is that the objects on the lower level cannot exist without objects on the higher level. These hierarchies exist within HPSN and are removed during the pre-network flattening process (see Section 6).

**4) Extraction of the relations between users and objects** – relation between a given user $a$ and object $X$ exists if user $a$ performed some activities towards object $X$, e.g. user $a$ commented photo $X$. The type of activity that user performed towards an object is assigned to each relationship.

The concept of HPSN is presented in Fig. 4 where the hierarchy between objects has three levels ($Z$, $V$ and $Y$), at each level some objects exist (e.g. at the level $Y$ objects *OB Y1* and *OB Y2*) with which users ($a$, $b$, $c$) are in different types of relations ($Z$, $V$ and $Y$ roles), i.e. users performed some activities towards these objects.

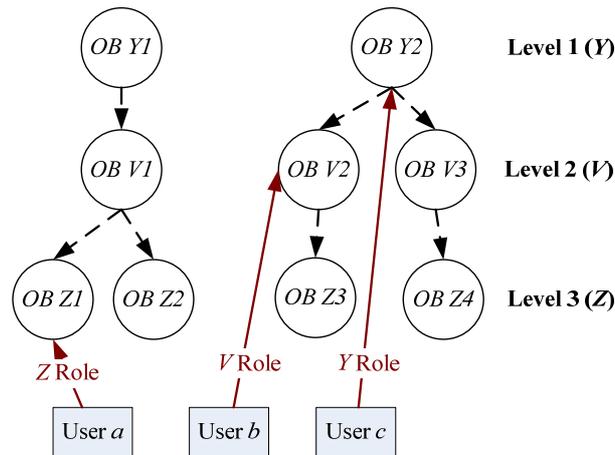

**Fig. 4 The concept of HPSN**

## 6. Flat Pre-Social Network

The flattening process aims at removing relationships between objects (hierarchies). A consequence of this process is that based on the knowledge about existing hierarchies both new user roles and relationships between users are created. The transformation from the pre-social network where the hierarchies between objects exist (HPSN) into the flat pre-social network (FPSN) without hierarchies will be performed in the following steps:

a) The operator chooses the level in the hierarchy to which the flattening process will be performed – the end level. Note that after each flattening process, the only object type in FPSN will be the one that is on the end level selected by the operator and all users will be in relations only towards these objects.
b) If there exist levels that are lower in the hierarchy than the end level (Fig. 5) then for those levels the **bottom-top** approach is used, i.e.:
  (i) Relationships between people and objects existing on the hierarchy levels that are below the end level (relation User *a* – OB Z1 and User *b* – OB V2 in Fig. 5) are changed. The relation between a user and an object from the lower lever is moved to the upper level by:
   - identification of an object on the upper level that is 'a father' of the object from the lower level ('child');
   - creation of a new relation between the user and 'the father' object;
   - name of the relation between user and 'father' object is created by adding to the name of the relation user – 'child' the word that denotes the movement from the lower level. For example, in Fig. 5: the relation user *a*–OB Z1 ('child') has a name: *Z Role* and the name of the new relation User *a*–OB V1 ('father') is *ZV Role*;
   - deletion of the relation between the user and the 'child object from the lower level.
     NOTE: This process is repeated for other upper levels until the end level is reached (Fig. 5).
  (ii) Relationships between people and objects existing at the end level remain unchanged (relation User *c* –OB Y2).

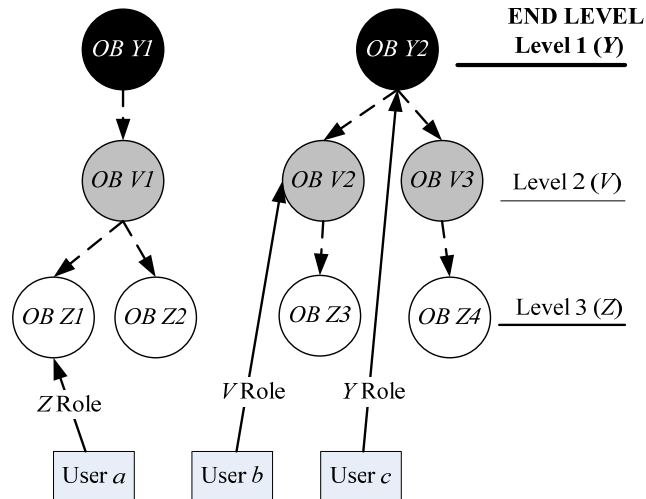

**Fig. 5 Relations between users and objects in the hierarchical pre-social network HPSN**

The final FPSN presented in Fig. 6 is an outcome of the bottom-up approach where the HPSN from Fig.5 is flatten to level 1 (*Y*).

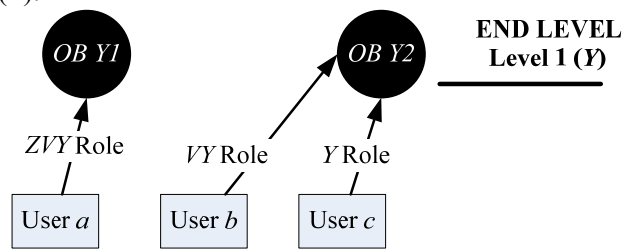

**Fig. 6 Relations between users and objects in the final FPSN after the bottom-top flattening process**

c) If there exist levels that are upper in the hierarchy than the end level (Fig. 7) then the **top-bottom** approach is applied for these levels, i.e.:
  (i) Relationships between people and objects existing on the hierarchy levels that are above the end level (relation User *b*–OB V2 and User *c*–OB Y2 in Fig. 7) are changed. The relation between user and object from the upper lever is moved to the lower level by:
  - identification of all objects on the lower level that are 'children' of an object form the upper level ('father'),
  - creation of the relation between the user and all 'child objects',
  - name of the relation between the user and 'child object' is created by adding to the origin name information about the 'child object'. For the example in Fig. 7: in the relation User *b*–OB V2 ('father') the relation name was: *V* Role and the new relation User *b*–OB Z3 ('child') will have the name: *VZ* Role,
  - deletion of the relation between the user and 'father object' on the upper level.
     NOTE: This process is repeated until the end level is reached (Fig. 8.).
  (ii) Relationships between people and objects existing on the end level remain unchanged.
  An example of top-bottom approach is presented below where the HPSN from Fig.7 is flatten to level 3 (*Z*), i.e. to FPSN in Fig. 8.

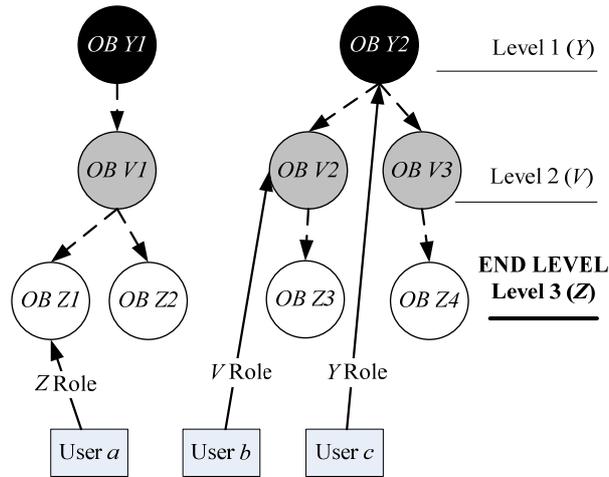
**Fig. 7 Relations between users and objects in the hierarchical pre-social network HPSN**

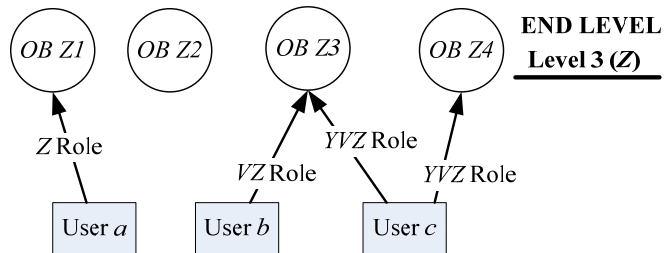
**Fig. 8 Relations between users and objects in the final FPSN after the top-bottom flattening process**

The goal of the flattening process is to facilitate the extraction of the unified structure that represents the social connections between pairs of users from user activity data and relations between objects. New types of user roles can be identified during the flattening process, e.g. *VZ* or *YVZ* Role in Fig. 8. The newly obtained knowledge about these roles gives an opportunity to investigate the complex profile of user relationships in more detail and in consequence enables their more comprehensive analysis.

## 7. Social Network

The flat pre-social network structure (FPSN) is used to extract the social network (SN) where the relations user-object from FPSN no longer exist. These connections are converted into direct relations between users in SN. The process consists of the following steps:
a) Extraction of SN layers based on the type of the users' roles towards objects. Each network layer consists of users and their connections. In a single layer network, there exist object-based relationships of only one type either with equal or different roles (see Section 3).
b) The operator chooses which social network layers need to be created.
c) The operator chooses the SN model:
  - **n-graph** – each layer in the multi-layered social network SN is represented by separate social network.
  - **multi-graph** – all layers are represented by a single social network and different layers are distinguished by different colours of edges (or another labelling mechanism is used).
d) Extraction of relations *user_from–user_to* by calculation of the relationship strengths and colours (labels) between SN nodes (users) using activity data stored in FPSN. There are many possible formulas for calculating the relationship strength. Most of them are based on the normalized quantity of shared user activities towards objects in FPSN (for some of the examples please see [12, 25]).

Social network SN created from FPSN in Fig. 8 according the process described above is presented in Fig. 9.

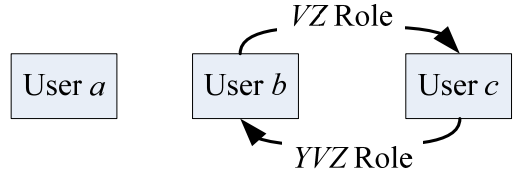

Fig. 9 Social network created from FPSN presented in Fig. 8

Depending on the goal of analysis, the strength of a relationship can be a static measure calculated based on all available data and taking into consideration the number of activities of a given type. On the other hand, we can take into account time factor and split the data according to the time when activities occurred. In the latter case, the whole period from which the data comes from is divided into time frames and the relationship strength is calculated for each slot separately. The time frame can be created using two approaches:

(i) **Sliding window** – a user defines the length of the time window (e.g. $t_l$) and the time interval that is used to move the window (e.g. $t_i$). In order to extract time frames the whole period of the length $t_l$ is moved by $t_i$. In consequence the entire dataset is divided into partly overlapping frames. Note that both time window and time interval need to be specified in a way that the period from the start date to end date should be completely covered.

(ii) **Equal, separate periods** – a user sets the number of periods, e.g. $k$, and then the data is divided into $k$ separate, equal periods according to the dates of activity. This is equivalent to the situation (i) where $t_l$ equals to $t_i$.

After the time windows are created, the weight is assigned to each of them. Usually, the recent periods are more important than the previous ones and because of that greater weight is assigned to those recent time windows .

In this paper we present how to calculate the static version of relationship strength for two different types of relations: (i) object-based relations with equal roles and (ii) object-based relations with different roles (see Fig. 2).

a) **Object-based relationships with equal roles**

The object-based relationship with equal roles denotes a connection in which two users are related to each other through the object and their roles towards this object are the same, see Section 3. Note that, the same formula is used in order to calculate the connection strength between user $x$ and user $y$ who (i) have commented at least one common object, (ii) have marked at least one common object as their favourite, (iii) are co-authors of an object, (iv) are in the same group or forum, (v) have used the same keywords to describe objects. In all of these relations, there is an object on which both users perform specific activity. To calculate the static strength of the relationship the following formula may be applied:

$$s_{xy}^a = \frac{n_{xy}^a}{n_x^a}, x \neq y , \qquad (1)$$

where:

$a$ – type of activity that is performed by users towards an object, e.g.: membership to a group/forum, utilization of a tag to describe objects, co-authorship of an object, commenting an object, etc.

$n_{xy}^a$ – number of common $a$ activities for users $x$ and $y$ performed together, e.g. number of groups/forums to which both users $x$ and $y$ belong, the number of tags that both users $x$ and $y$ use commonly or the number of objects that were co-authored by both users $x$ and $y$, etc.

$n_x^a$ – number of a given *a* activity for user *x*, e.g. the number of groups/forums to which user *x* belongs, the number of tags used by user *x* or the number of objects authored by user *x*, etc.

Let us consider the situation in which users of multimedia sharing system utilise tags to described different multimedia content. In this case an object is *a tag* and relationship between two users is created when they utilise some common tags. Let us assume that the data obtained from the system contains the following information: user *x* utilised 20 identical tags as user *y* and user *x* used 60 tags in total. Then relation strength from user *x* to *y* is calculated as follows:

$$s_{ij}^y = \frac{n_{ij}^y}{n_i^y} = \frac{20}{60} = \frac{1}{3}.$$

**b) Object-based relationships with different roles**

The object-based relation with different roles denotes a connection in which two users are related to each other through the object and their roles towards this object are different (see Section 3). For example, one user can comment in a forum in which another user is a moderator and the relationship between users is: moderator-commentator. Thus, in the case of calculating the relations strength, we will refer to the relations *activity_type_a-activity_type_b*.

Considering the relation *activity_type_a-activity_type_b* its strength will be calculated as follows:

$$s_{xy}^{ab} = \frac{n_{xy}^{ab}}{n_x^{ab}}, x \neq y, a \neq b , \qquad (2)$$

where

*a* – denotes 1st activity type;

*b* – denotes 2nd activity type (different that *a*);

$n_{xy}^{ab}$ – number of activities *a* performed by user *x* towards objects for which user *y* performed activity *b*;

$n_x^{ab}$ – total number of activities of type *a* performed by user *x* towards objects for which any other users performed activity *b*.

As the example, let us consider the case where a user adds to the list of favourites an object authored by another user. A relationship between two users is created when one user adds to its favourites an object authored by another person. Assume that the following data is available in the system: user *x* added to favourites 20 objects authored by user *y*. User *x* added to favourites 60 objects in total. The objects of user *y* were added to favourites by others 30 times in total. Moreover, *a* means the activity 'authored by' and *b* – the activity 'added to favourite by'.

The relation strengths are calculated as follows:

from user *x* to *y*: $s_{xy}^{ba} = \frac{n_{xy}^{ba}}{n_x^{ba}} = \frac{20}{60} = \frac{1}{3}$

from user *y* to *x*: $s_{yx}^{ab} = \frac{n_{yx}^{ab}}{n_y^{ab}} = \frac{20}{30} = \frac{2}{3}$

## 8. Example of Flattening Process

One of the examples where hierarchies between objects exist is internet forum where people can create their own topics that contain posts added by users. The hierarchy between objects within a forum and the activities that can be performed towards these objects are presented in Fig. 10.

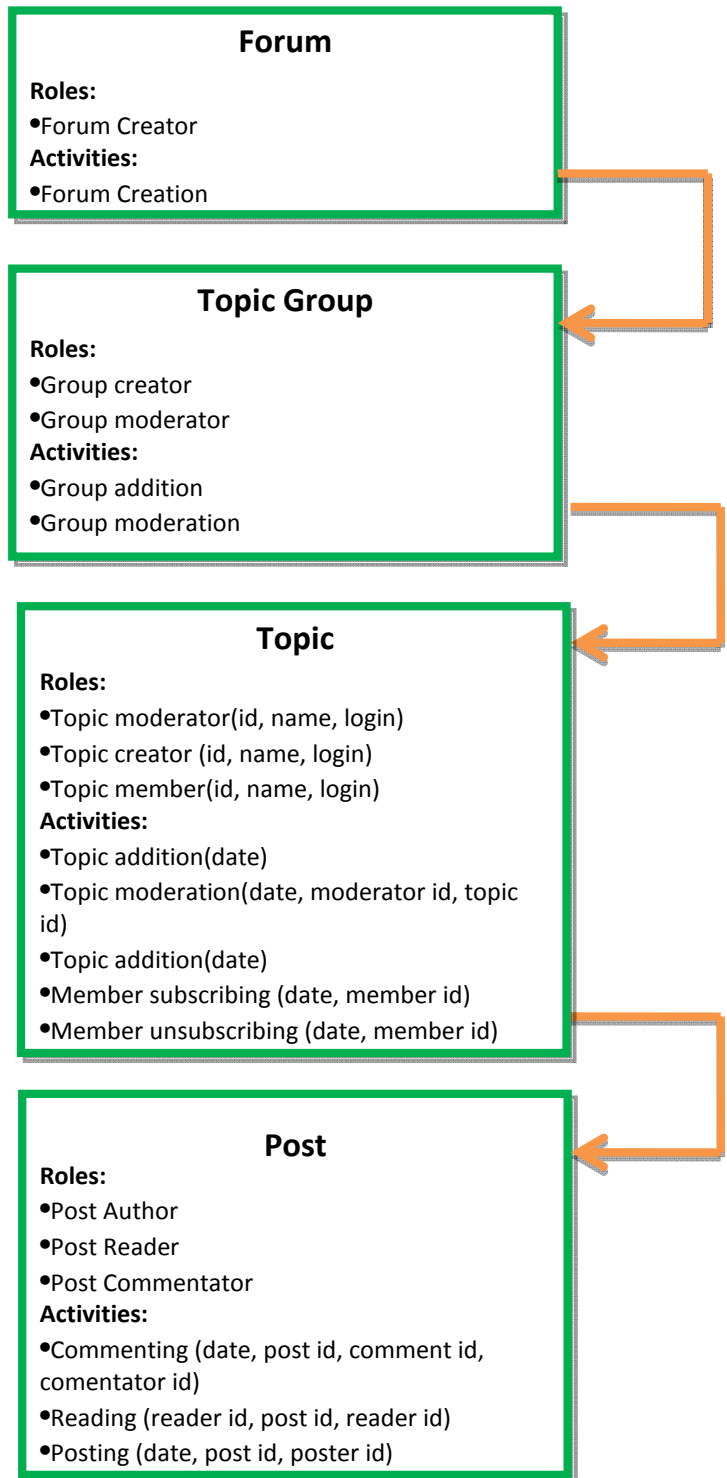

**Fig. 10 The hierarchy of objects in the forum**

The hierarchy that will be used for our case study is: forum – topic – post. Both relationships between objects and between users and objects in the exemplary hierarchical pre-network are presented in Fig. 11. In order to create the flat pre-social network (Fig. 12) we perform the flattening process after which the relationships between objects will be removed but at the same time on other levels new relationships between users and objects will be created.

Two types of flattening process can be considered: bottom-top and top-bottom (see Section 6). We present here the bottom-top flattening process in which the relationships will be moved to the highest, forum, level, i.e. the forum will be the final level.

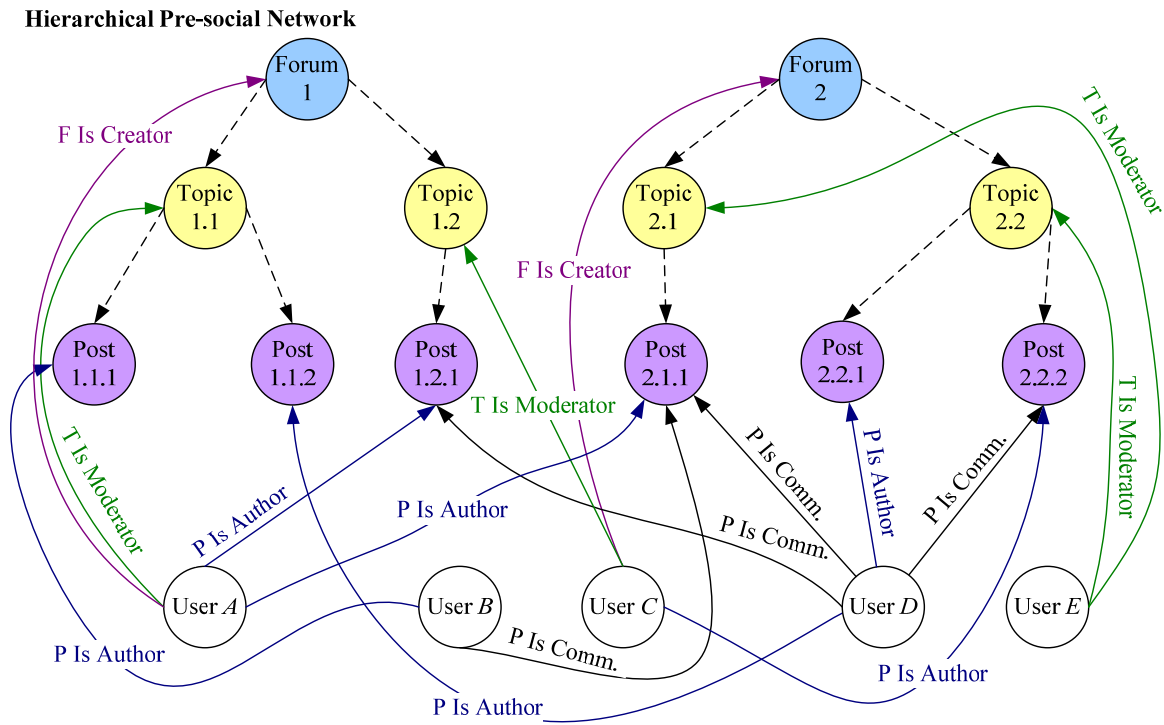

**Fig. 11 Relations between users and objects together with users' roles towards the objects**

The bottom-top approach applied to HPSN from Fig. 11 results in several new relationships and roles (Fig. 12). Some examples are enumerated below:
- User *A* is an author of post 1.2.1 in the topic 1.2 at the forum 1 and also of post 2.1.1 in the topic 2.1 at forum 2. Then a new relation between User *A* and forum is created: *PTF Is Author (PTF - PostTopicForum* – a new name of relation, see Section 6), i.e. User *A* authored at least one post in a topic that is in a given forum. A similar approach is applied to User *B*, who is an author of post 1.1.1 in the topic 1.1 and to User *C* - an author of post 2.2.2 in the topic 2.2. Moreover, User *D* is an author of posts 1.1.2 and 2.2.1. This is flattening (movement) of authorship activities on posts (role *P Is Author*) to the top level – forum.
- User *B* is a commentator of post 2.1.1. A new relation between User *B* and the forum is created: *PTF Is Commentator*. The same is done for User *D,* an commentator of posts 1.2.1, 2.1.1 and 2.2.2. In this way, the role *P Is Commentator* is moved to the forum level.
- User *A* is a moderator of topic 1.1 at the forum – then a new relation between User *A* and the forum is create*d – TF Is Moderator (TF - TopicForum* – a new name of relation, see Section 6). The similar method is utilized for User *C* - the moderator of topic 1.2 and User *E* – the moderator of topics 2.1 and 2.2. This is flattening of roles from the topic level to the final forum level.
- *User A* is a creator of the forum 1 and User *C* is a creator of forum 2 – then the existing relationship stays unchanged: *F Is Creator*.

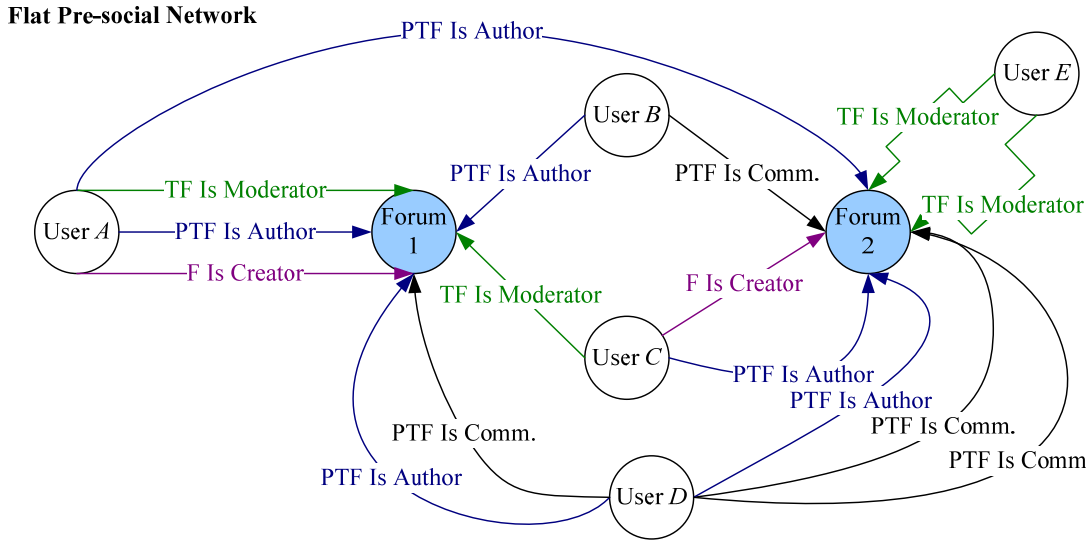

**Fig. 12 Flat pre-social network (FPSN) – the result of flattening process applied to HPSN from Fig. 11**

Two different layers of the final multi-layered SN derived from the flat pre-social network FPSN (Fig. 12) are presented in Fig. 13 and Fig. 14. For instance, the relation between two users exists if one user was the moderator of the post that was commented by another user, Fig. 13, e.g. User *A* moderates the post commented by User *D* so there is a relation *moderator-commentator* from *A* to *D* in the final social network SN. This relation is an object-based relationship with different roles. In Fig. 14, another layer *PTF Is Author – PTF Is Author is* presented. This is a layer in which the extracted relationships are object-based with equal roles. The relationships' strengths presented on both figures are calculated using formulas from Section 7.

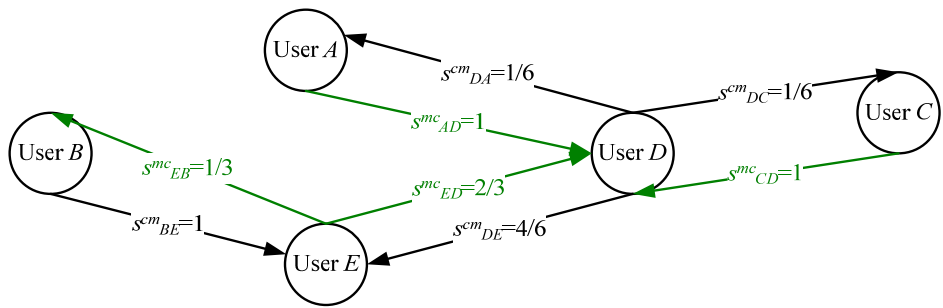

**Fig. 13 Example of the *PTF Is Comm. – MF Is Moderator* layer of the final SN extracted from the flat pre-social network FPSN (Fig. 12).**

Note that User *D* and User *E* are connected only because of flattening process (Fig. 13). The same is with User *E* and User *B* (Fig. 13) and User *A* and User *C* (Fig 14). Any of these relationships would be revealed if the flattening process did not take place.

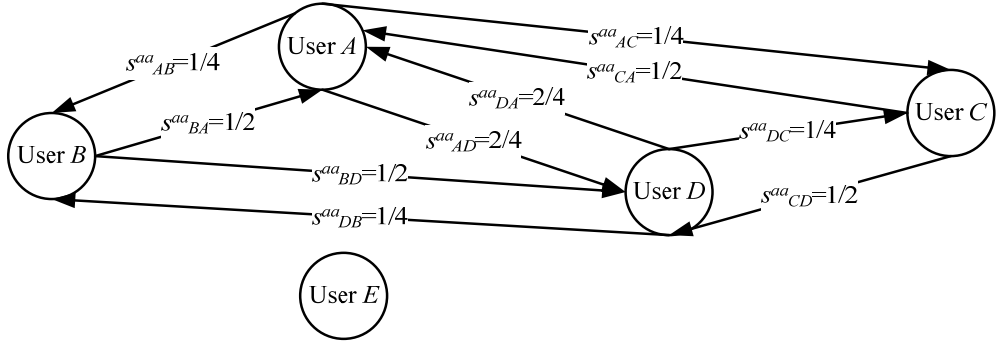

**Fig. 14** Example the *PTF Author – PTF Author* layer of the final SN extracted from the flat pre-social network FPSN (Fig. 12).

## 9. Experiments

The real-world dataset used for experiments was obtained from the social web site *extradom.pl*. The analysed dataset covers the period from 21-08-2008 to 08-01-2010. Before the hierarchical pre-social network HPSN was created, the dataset was cleansed and validated. Several rules were applied in the cleansing phase. Two most important ones were:
- each object must have creation date;
- each object must be assigned to its creator.

| Object type | No. of objects |
|---|---|
| forum | 1 |
| topic group | 692 |
| topic | 2,336 |
| post | 13,272 |
| comment | 49 |

**Tab. 1. Objects quantity in the experimental dataset**

There were 104,625 users registered in the portal, but only 4.25% (4,404) of them were active on forum. The number of different types of objects in the forum is shown in Tab. 1. One or more activity type were identified for each object type. Objects with activities that were performed towards them are shown in Tab. 2.

| Object type | Activity type | No. of activities |
|---|---|---|
| forum | forum creation activity | 1 |
| topics group | topic group addition | 692 |
| topics group | topic group moderation | 0 |
| topic | topic addition | 2,336 |
| topic | topic member subscribing | 5,788 |
| post | post reading | 0 |
| post | post authoring | 13,272 |
| comment | post commenting | 49 |

**Tab. 2. Activities assigned to object types**

As we can see in Tab. 2, there are two activity types (*topic group moderation* and *post reading*) that are not present in the dataset used for experiments and thus are not included in the further analyses. Additional assumptions that have been made and which helped to detect some of activities are:
- a user who creates the first post in the topic will be treated as a topic creator;
- creation of the first topic in the group is simultaneously treated as the creation of the entire group;
- users who create their first post in the topic will be automatically subscribed to this topic.

Tab. 3 summarizes the profile of the hierarchical pre-social network (HPSN) that was created from the *extradom.pl* dataset.

| Activity type | No. of activities | No. of users with the activity | Percentage of total users with a given activity |
|---|---|---|---|
| forum creation activity | 1 | 1 | 0.02% |
| topic group addition | 692 | 266 | 6.04% |
| topic addition | 2,336 | 1,464 | 33.24% |
| topic member subscribing | 5,788 | 4,359 | 98.98% |
| post posting | 13,272 | 4,359 | 98.98% |
| post commenting | 49 | 39 | 0.89% |

**Tab. 3. Activities in the hierarchical pre-social network**

Three distinct flattening processes (see Section 6) with three separate final object levels: topics group, topic and post have been applied to the hierarchical pre-social network HPSN. Some of the activities were multiplied after the flattening process (for detailed statistics please see Tab. 4). Such situation takes place when the hierarchical pre-social network is flattened to the object type, which is not the highest level in the hierarchy. For example, when HPSN was flattened to the topic groups (level 2 in the hierarchy, see Fig. 3), the forum creation activity was multiplied by 692 because there were distinct 692 groups in the dataset.

| Activity type | No. of activities before flattening (in HPSN) | New | Final objects: topic groups | New | Final objects: topics | New | Final objects: posts |
|---|---|---|---|---|---|---|---|
| forum creation activity | 1 | + | 692 | + | 2336 | + | 1,3272 |
| topic group addition | 692 | | 692 | + | 2336 | + | 13,272 |
| topic addition | 2,336 | | 2,336 | | 2336 | + | 13,272 |
| topic member subscribing | 5,788 | | 5,788 | | 5788 | | 793,245 |
| post authoring | 13,272 | | 13,272 | | 13,272 | | 13,272 |
| post commenting | 49 | | 49 | | 49 | | 49 |

**Table 4. Number of activities before and after flattening; '+' denotes activates which were created during a given flattening process**

Once the flattening process has been accomplished, the separate layers in the multi-layered social network were identified and relationships between users within these layers were extracted (see Section 7). Both layers and number of distinct relationships existing within each layer are presented in Tab. 5.

| Layer | New | Moved | Topic groups | Moved | Topics | Topics / Topic groups | Moved | Posts | Posts / Topics |
|---|---|---|---|---|---|---|---|---|---|
| forum creation activity - group addition | + | | 671 | | 2,252 | 3.36 | | 12,922 | 5.74 |
| forum creation activity - topic addition | + | | 2,313 | | 2,313 | 1.00 | | 13,157 | 5.69 |
| forum creation activity - member subscribing | + | | 5,760 | | 111,544 | 19.37 | | 792,732 | 7.11 |
| forum creation activity - post authoring | + | | 13,233 | | 13,233 | 1.00 | | 13,233 | 1.00 |
| forum creation activity - commenting | + | | 49 | | 49 | 1.00 | | 49 | 1.00 |
| group addition - topic addition | + | | 1,614 | | 1,614 | 1.00 | | 8,821 | 5.47 |
| group addition - member subscribing | + | | 5,116 | | 109,377 | 21.38 | | 780,312 | 7.13 |
| group addition - post authoring | + | | 12,206 | | 12,206 | 1.00 | | 12,206 | 1.00 |
| group addition - commenting | + | | 49 | | 49 | 1.00 | | 49 | 1.00 |
| topic addition - member subscribing | | + | 109,376 | | 109,376 | 1.00 | + | 780,445 | 7.14 |
| topic addition - post authoring | + | | 302,961 | | 9,849 | 0.03 | | 9,849 | 1.00 |
| topic addition - commenting | + | | 1,060 | | 44 | 0.04 | | 44 | 1.00 |
| member subscribing - member subscribing | | + | 289,740 | + | 12,342,690 | 42.60 | + | 83,495,278 | 6.76 |
| member subscribing - post authoring | + | | 779,973 | | 779,973 | 1,00 | | 779,973 | 1.00 |
| member subscribing - commenting | + | | 2,903 | | 2,903 | 1.00 | | 2,903 | 1.00 |
| post authoring - post authoring | + | | 2,449,226 | | 376,978 | 0.15 | | 0 | 0 |
| post authoring - commenting | | + | 9,340 | + | 1,703 | 0.18 | | 44 | 0.03 |
| Sum: | | | 3,985,590 | | 13,876,153 | 3.48 | | 86,702,017 | 6.25 |
| | | | New: 90% | | New: 10% | | | New: 3% | |
| | | | Moved: 10% | | Moved: 90% | | | Moved: 97% | |

**Table 5. Layers in SN and their profile for three different flattening processes**

As a result of the flattening process 14 new layers in the multi-layered social network were created. Additionally, percentages of new user relationships in SN are: 90% in the case of topic groups as the final object, 10% for topics, and 3% for posts (see Tab. 5, Fig. 15). Note, that these new relationships would not be visible without the flattening process. It means that the method of pre-processing with flattening of object relations reveals completely new knowledge about the complexity of connections between people.

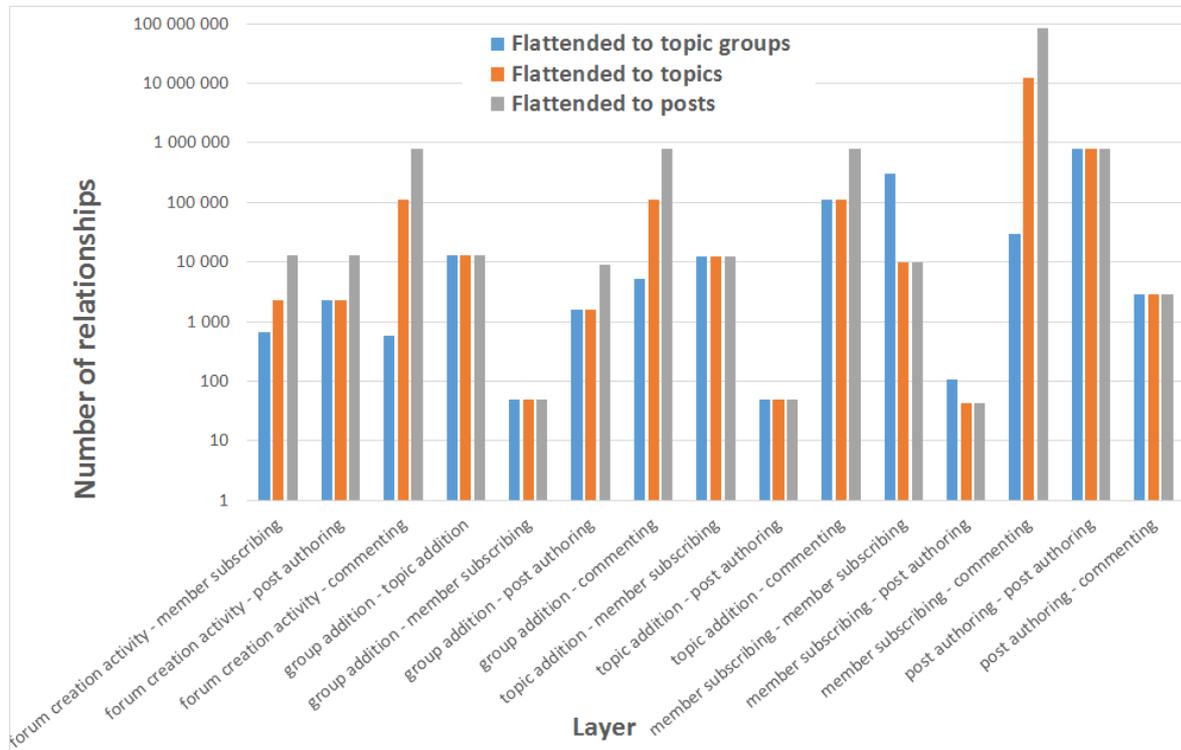

**Fig. 15. No. of activities (relationships) within created layers for three different flattening processes**

## 10. Conclusions

The wide variety and availability of Web 2.0 systems, where users can interact with each other and perform different types of activities, give us an opportunity, by analysing the large-scale data gathered in these systems, to better understand human social behaviour. A very interesting research problem is to investigate social connections that emerge between people based on their shared activities. However, the extraction of these relations is not a trivial task. The main reason is that user behaviour in such systems is often very complex due to the variety of available services and functionalities. As presented in the paper, people can perform different activities towards different objects. Additional challenge is that the relationships existing between these objects can form a hierarchical structure. In this paper, we propose the process to extract the multi-layered social network from the data about both user behaviours and relations between objects. The whole method consists of three main phases: (i) extraction of the hierarchical pre-social network HPSN, (ii) creation of the flat pre-social network PFSN and finally (iii) creation of the multi-layered social network SN. We believe that such systematic approach to the problems is necessary to be able to cope with the massive volume of data being generated by social-based systems every day. Moreover, the proposed process is generic and robust in a way that it is able to accommodate new ways of interactions between users.

The new flattening concept enables to discover in the multi-layered social network new layers with new types of relationships which otherwise would not be available for analysis. It is possible due to the presented above process in which the object hierarchy is removed. Thus, the new method of pre-processing enables to reveal new information, which is invisible in the regular analysis of user activities and this in turn opens new possibilities for network analysis. The experiments confirmed that some new types of relations between users can be extracted in the flattening process. This enables a deeper insight into analysis of multi-layered social networks as more information is included in the final network structure.

**Acknowledgments**

The authors are indebted to Elżbieta Kukla, PhD, Tomasz Filipowski, MSc for their valuable discussion. The work was partially supported by Fellowship co-Financed by European Union within European Social Fund, by The European Commission under the 7th Framework Programme, Coordination and Support Action, Grant Agreement Number 316097, ENGINE - European research centre of Network intelliGence for INnovation Enhancement http://engine.pwr.wroc.pl/ and by The National Science Centre, the decision no. DEC-2013/09/B/ST6/02317.

## 11. References

[1] Abraham I., Chechik S., Kempe D., Slivkins A.: Low-distortion Inference of Latent Similarities from a Multiplex Social Network. The Twenty-Fourth Annual ACM-SIAM Symposium on Discrete Algorithms, SODA 2013, SIAM 2013, pp. 1853-1872.
[2] Adamic L.A., Adar E.: Friends and Neighbors on the Web, Social Networks, 25(3), 2003, pp. 211-230.
[3] Agarwal, N., Galan, M., Liu, H., Subramanya S.: WisColl: Collective Wisdom based Blog Clustering, Information Sciences, Vol. 180, Issue 1, 2010, pp. 39-61.
[4] Baptista M.S., Grebogi C., Köberle R.: Dynamically multilayered visual system of the multifractal fly. Phys Rev Lett. 97(17) 2006, 178102.
[5] Barrat A., Barthelemy M., Vespignani A.: Dynamical Processes on Complex Networks, Cambridge University Press, UK, 2008.
[6] Berlingerio M., Coscia M., Giannotti F., Monreale A., Pedreschi D.: The pursuit of hubbiness: Analysis of hubs in large multidimensional networks. J. Comput. Science 2(3), 2011, pp. 223-237.
[7] Blondel V.D., Guillaume J.-L., Lambiotte R., Lefebvre E.: Fast unfolding of communities in large networks, J. Stat. Mech., P10008, 2008.
[8] Bródka P., Kazienko P.: Multi-layered social networks. Encyclopedia of Social Network Analysis and Mining. Springer, Summer 2014, in press.
[9] Bródka P., Kazienko P., Musiał K., Skibicki K.: Analysis of Neighbourhoods in Multi-layered Dynamic Social Networks. International Journal of Computational Intelligence Systems, Vol. 5, No. 3, June, 2012, pp. 582-596.
[10] Bródka P., Skibicki K., Kazienko P., Musiał K.: A Degree Centrality in Multi-layered Social Network. CASoN 2011, The International Conference on Computational Aspects of Social Networks, IEEE Computer Society, 2011, pp. 237-242.
[11] Bródka P., Stawiak P., Kazienko P.: Shortest Path Discovery in the Multi-layered Social Network. ASONAM 2011, The 2011 International Conference on Advances in Social Network Analysis and Mining, IEEE Computer Society, 2011, pp. 497-501.
[12] Caldarelli G., Vespignani A. (eds.): Large Scale Structure and Dynamics of Complex Networks, From Information Technology to Finance and Natural Science, Complex Systems and Interdisciplinary Science, vol. 2, World Scientific Publishing Co. Pte. Ltd., 2007.
[13] Capocci A., Servedio V., Colaiori, F., Buriol L., Donato D., Leonardi S., Caldarelli G.: Preferential attachment in the growth of social networks: The internet encyclopedia Wikipedia, Physical Review E, vol. 74, Issue 3, id. 036116, 2006.
[14] Cheng X., Dale C., Liu J.: Statistics and social networking of YouTube videos. 16th International Workshop on Quality of Service, IWQoS 2008, IEEE, 2008, pp. 229-238.
[15] Contractor N.S., Monge P.R., Leonardi P.M.: Multidimensional networks and the dynamics of sociomateriality: Bringing technology inside the network. International Journal of Communication. 5(1), 2011, pp. 682-720.

[16] Coscia M., Giannotti F., Pedreschi D.: A Classification for Community Discovery Methods in Complex Networks. 2012, arXiv:1206.3552 [cs.SI].
[17] De Domenico M., Sole-Ribalta A., Cozzo E., Kivela M., Moreno Y., Porter M.A., Gomez S., Arenas A.: Mathematical formulation of multi-layer networks, Physical Review X, 3, 2013, 041022.
[18] Ellison, N.B., Steinfield, C., Lampe, C.: The benefits of Facebook "friends:" Social capital and college students' use of online social network sites. Journal of Computer-Mediated Communication, 12(4), article 1. http://jcmc.indiana.edu/vol12/issue4/ellison.html, 2007.
[19] Girvan M., Newman M.E.J., Community structure in social and biological networks, Proc. Natl. Acad. Sci., USA, 99 (12) (2002), pp. 7821–7826.
[20] Gomez S., Diaz-Guilera A., Gomez-Gardenes J., Perez-Vicente C.J., Moreno Y., Arenas A.: Diffusion dynamics on multiplex networks, Physical Review Letters, 110, 2013, 028701.
[21] Hanneman, R., Riddle, M.: Introduction to social network methods, online textbook, Riverside, CA: University of California, 2005, http://faculty.ucr.edu/~hanneman/nettext/.
[22] Huberman B., Romero D., Wu F.: Social networks that matter: Twitter under the microscope. First Monday, 2009, pp 1-5 (arXiv:0812.1045v1).
[23] Kazienko P., Bródka P., Musial K.: Individual Neighbourhood Exploration in Complex Multi-layered Social Network. WI-IAT 2010, 2010 IEEE/WIC/ACM International Conference on Web Intelligence and Intelligent Agent Technology, CISWSN 2010, The Third Workshop on Collective Intelligence in Semantic Web and Social Networks workshop, WI-IAT 2010 Workshops proceedings, IEEE Computer Society Press, 2010, pp. 5-8.
[24] Kazienko P.,Bródka P., Musial K., Gaworecki J.: Multi-layered Social Network Creation Based on Bibliographic Data. SocialCom-10, The Second IEEE International Conference on Social Computing, SIN-10 Symposium on Social Intelligence and Networking, IEEE Computer Society Press, 2010, pp. 407-412.
[25] Kazienko P., Musial K., Kajdanowicz T.: Multidimensional Social Network in the Social Recommender System. IEEE Transactions on Systems, Man, and Cybernetics - Part A: Systems and Humans, Vol. 41, Issue 4, 2011, pp. 746-759.
[26] Kazienko P., Ruta D., Bródka P.: The Impact of Customer Churn on Social Value Dynamics. International Journal of Virtual Communities and Social Networking, 1(3), July-September 2009, pp. 60-72.
[27] Lee S., Monge P.: The Coevolution of Multiplex Communication Networks in Organizational Communities. Journal of Communication, Volume 61, Issue 4, August 2011, pp. 758-779.
[28] Lomi A., Lusher D., Pattison P. E., Robins G.: Inter-organizational hierarchies, social networks, and identities in multi-unit organizations, American Sociological Association, TBA, New York, 2007, http://www.allacademic.com/meta/p183413_index.html.
[29] Michalski R., Kazienko P., Jankowski J.: Convince a Dozen More and Succeed - The Influence in Multi-layered Social Networks. The Second Workshop on Complex Networks and their Applications at SITIS 2013 - The 9th International Conference on Signal Image Technology & Internet based Systems, IEEE Computer Society, pp. 499-505.
[30] Mika P.: Flink: Semantic Web technology for the extraction and analysis of social networks, Web Semantics: Science, Services and Agents on the World Wide Web, Vol. 3, Issues 2-3, Selected Papers from the International Semantic Web Conference, 2004 - ISWC, 2004, October 2005, pp. 211-223, ISSN 1570-8268.
[31] Matsuo Y., Mori J., Hamasaki M., Nishimura T., Takeda H., Hasida K., Ishizuka M.: POLYPHONET: An advanced social network extraction system from the Web. Journal of Web Semantics: Science, Services and Agents on the World Wide Web, Volume 5, Issue 4, December 2007, pp. 262-278, ISSN 1570-8268.
[32] Musiał K., Kazienko P.: Social Networks on the Internet. World Wide Web, Vol.16, No.1, 2013, pp. 31-72.
[33] Pasqual M.C., de Weck O.L.: Multilayer network model for analysis and management of change propagation. Research in Engineering Design, Volume 23, Issue 4, October 2012, pp 305-328.
[34] Sahneh F.D., Scoglio C., Van Mieghem P.: Generalized Epidemic Mean-Field Model for Spreading Processes Over Multilayer Complex Networks. EEE/ACM Transactions on Networking, Vol. 21, No. 5, October 2013, pp. 1609-1620.
[35] Scott J.: Social Network Analysis: A Handbook, SAGE Publications, London, UK, 2000.
[36] Strogatz S. H.: Exploring complex networks, Nature, 410(6825), March 1998, pp. 268–276.
[37] Traud A.L., Kelsic E.D., Mucha P.J., Porter M.A.: Community structure in online collegiate social networks, eprint arXiv:0809.0690.

[38] Tyler J.R., Wilkinson D.M., Huberman B.A.: Email as spectroscopy: Automated discovery of community structure within organizations, in: Communities and Technologies, Kluwer, B.V., Deventer, The Netherlands, 2003, pp. 81-96.
[39] Wasserman, S., Faust, K.: Social network analysis: Methods and applications. Cambridge University Press, New York, 1994.
[40] Watts D.J., Strogatz S.: Collective dynamics of 'small-world' networks. Nature 393, June 1998, pp. 440-444.
[41] Zagenczyk T.J., Purvis R.L., Shoss M.K., Scott K.L., Cruz K.S.: Social Influence and Leader Perceptions: Multiplex Social Network Ties and Similarity in Leader–Member Exchange. Journal of Business and Psychology, 2013, DOI 10.1007/s10869-013-9332-7.
[42] Zignani M., Quadri C., Gaitto S., Rossi G.P.: Exploiting all phone media? A multidimensional network analysis of phone users' sociality. 2014, arXiv:1401.3126 [cs.SI].